\documentstyle[12pt,a4]{article}
\input epsf
\global\arraycolsep=2pt

\catcode`\@=11
\newcommand{\smalllineskip}{\baselineskip=12pt}
\newcommand{\fcaption}[1]{
        \refstepcounter{figure}
        \setbox\@tempboxa = \hbox{\small Fig.~\thefigure. #1}
        \ifdim \wd\@tempboxa > 5in
           {\begin{center}
        \parbox{5in}{\small\smalllineskip Fig.~\thefigure. #1}
            \end{center}}
        \else
             {\begin{center}
             {\small Fig.~\thefigure. #1}
              \end{center}}
        \fi}

\begin{document}

\begin{titlepage}

\begin{flushright}
BARI-TH/96-236\\
CERN-TH/96-125\\
hep-ph/9605274\\
May 1996
\end{flushright}

\vspace{1.0cm}

\begin{center}
\boldmath
\Large\bf Measuring $\alpha_s(Q^2)$ in $\tau$ Decays
\unboldmath
\end{center}

\vspace{1.0cm}

\begin{center}
Maria Girone\\
{\sl Dipartimento di Fisica, INFN Sezione di Bari, 70126 Bari,
Italy}\\
\vspace{0.3cm}
and\\
\vspace{0.3cm}
Matthias Neubert\\
{\sl Theory Division, CERN, CH-1211 Geneva 23, Switzerland}
\end{center}
\vspace{1.0cm}

\begin{abstract}
The decay rate of the $\tau$ lepton into hadrons of invariant mass
smaller than $Q\gg\Lambda_{\rm QCD}$ can be calculated in QCD using
the OPE. Using experimental data on the hadronic mass distribution,
the running coupling constant $\alpha_s(Q^2)$ is extracted in the
range $0.85~\mbox{GeV}<Q<m_\tau$, where its value changes by about a
factor~2. At $Q=m_\tau$, the result is $\alpha_s(m_\tau^2)=0.33\pm
0.03$, corresponding to $\alpha_s(m_Z^2)=0.119\pm 0.004$. The running
of the coupling constant is in excellent agreement with the QCD
prediction based on the three-loop $\beta$-function.
\end{abstract}

\vspace{1.0cm}

\begin{center}
\it To appear in the Proceedings of\\
Les Rencontres de Physique de la Vall\'ee d'Aoste\\
(La Thuile, Italy, March 1996)\\
and\\
Second Workshop on Continuous Advances in QCD\\
(Minneapolis, Minnesota, March 1996)
\end{center}

\end{titlepage}

\section{Introduction}

One of the most accurate methods to determine $\alpha_s$ in the
low-energy region is provided by the measurement of $R_\tau$, the
$\tau$ decay rate into hadrons normalized to the leptonic decay rate:
\begin{equation}
   R_\tau = {\Gamma(\tau\to\nu_\tau + \mbox{hadrons})
        \over\Gamma(\tau\to\nu_\tau\,e\,\bar\nu_e)} \,.
\end{equation}
$R_\tau$ can be calculated in QCD using the Operator Product
Expansion (OPE) \cite{SVZ,Rtau1}. The result is:
\begin{eqnarray}\label{Rtau}
   R_\tau &=& N_c\,\Big\{ 1 + \delta_{\rm pert}[\alpha_s(m_\tau^2)]
    + \delta_{\rm power} \Big\} \phantom{ \bigg[ } \nonumber\\
   &=& N_c\,\Bigg\{ 1 + {\alpha_s(m_\tau^2)\over\pi}
    + 5.202\,\bigg( {\alpha_s(m_\tau^2)\over\pi} \bigg)^2
    + 26.37\,\bigg( {\alpha_s(m_\tau^2)\over\pi} \bigg)^3
    + \dots \nonumber\\
   &&\phantom{ N_c\,\Bigg\{ 1 }
    - 8\,|V_{us}|^2\,{m_s^2\over m_\tau^2}
    + 32\pi^2\,{\langle m\bar\psi\psi\rangle\over m_\tau^4}
    - 2\,{\langle O_6\rangle\over m_\tau^6} + \dots \Bigg\} \,.
\end{eqnarray}
The non-perturbative power corrections in this expression are
proportional to the strange-quark mass, the quark condensate, and
higher-dimensional condensates. Because all contributions of
dimension less than six vanish in the chiral limit, the power
corrections are numerically small; using standard values of the QCD
parameters, one finds $\delta_{\rm power}=-(1.4\pm 0.5)\%$. This,
together with the fact that the perturbation series is known to third
order, make $R_\tau$ a good observable to measure $\alpha_s$.

\begin{figure}[ht]
   \epsfxsize=9cm
   \centerline{\epsffile{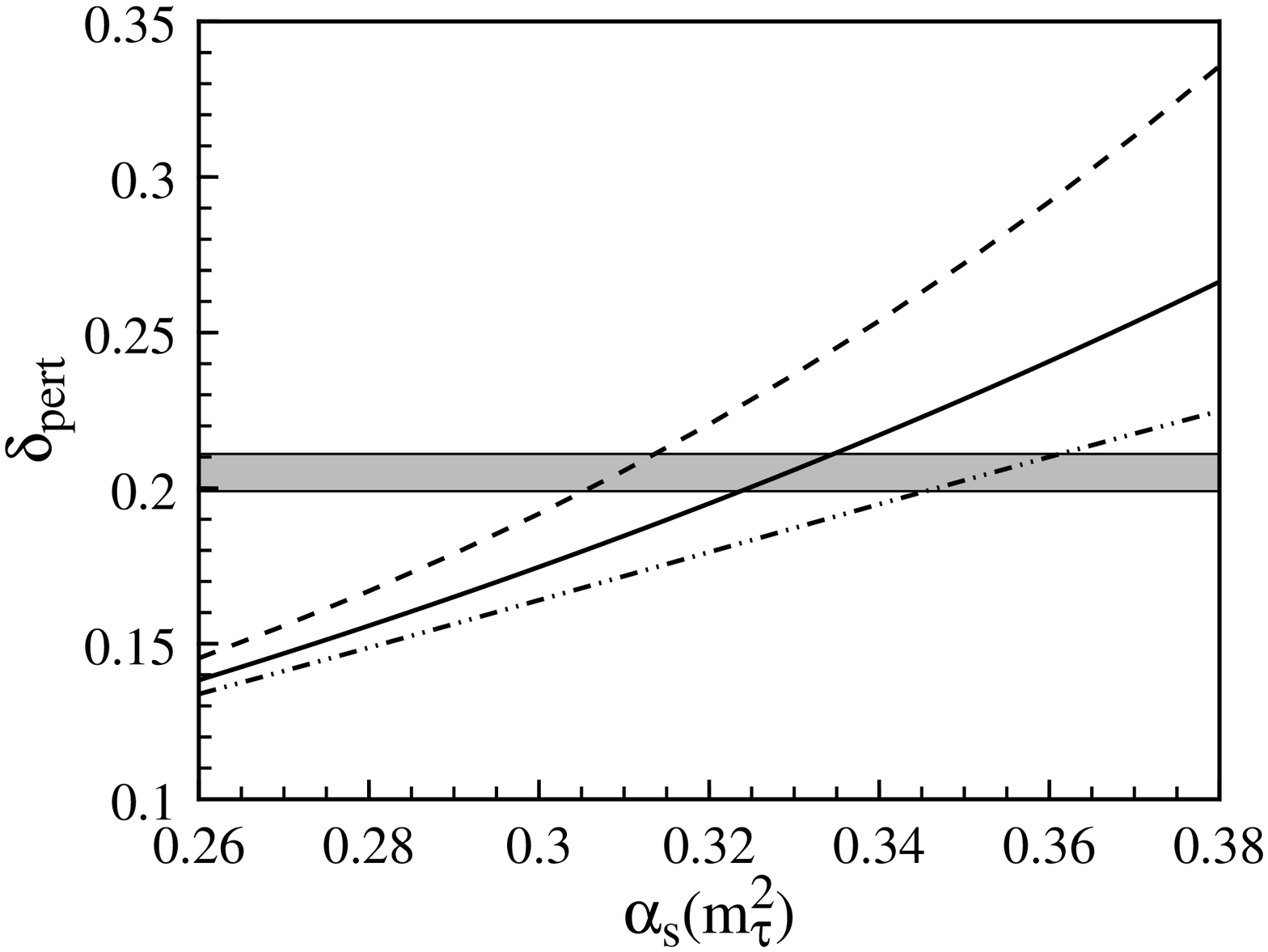}}
\fcaption{\label{fig:dpert}
Different perturbative approximations for the quantity $\delta_{\rm
pert}$: exact order-$\alpha_s^3$ result in the $\overline{\mbox{\sc
ms}}$ scheme (solid line), resummation of renormalon chains (dashed
line), resummation of Le~Diberder and Pich (dash-dotted line). The
experimental result (\protect\ref{dval}) is shown as a band.}
\end{figure}

Experimentally, $R_\tau$ is obtained from the relation
$R_\tau=1/B_e-1.97256$, where $B_e$ is the leptonic branching ratio.
Direct measurements give $B_e=(17.80\pm 0.06)\%$ \cite{Callot},
whereas using the $\tau$ lifetime, $\tau_\tau=(291.3\pm 1.6)$~fs
\cite{Wash}, we obtain $B_e=\tau_\tau/\tau_\mu\,(m_\tau/m_\mu)^5
=(17.84\pm 0.10)\%$. Averaging the two results gives $R_\tau=3.642\pm
0.010$, and taking into account small electroweak radiative
corrections not displayed in (\ref{Rtau}) we obtain
\begin{equation}\label{dval}
   \delta_{\rm pert}[\alpha_s(m_\tau^2)] = 0.205\pm 0.003_{\rm exp}
   \pm 0.005_{\rm th} \,.
\end{equation}
The dominant theoretical uncertainty in extracting $\alpha_s$ from
this measurement comes from the truncation of perturbation
theory \cite{Alta}, which induces an error of order $\alpha_s^4$.
This uncertainty can be estimated by considering some approximate
resummations of the perturbation series (starting at order
$\alpha_s^4$) and comparing them to the fixed-order calculation. The
resummation procedure of Le~Diberder and Pich \cite{LP} resums
certain ``large-$\pi^2$'' terms to all orders in perturbation theory.
Recently, another class of terms, the so-called renormalon
chains \cite{tHof}, have been investigated. These are the terms of
order $\beta_0^{n-1}\alpha_s^n$ in the perturbation series for
$\delta_{\rm pert}$, where $\beta_0$ is the first coefficient of the
QCD $\beta$-function. The resummation of such terms in the case of
$R_\tau$ has been discussed in Refs.~\cite{BBB,MN}. In
Fig.~\ref{fig:dpert}, we show the corresponding theoretical
predictions for $\delta_{\rm pert}$ as a function of
$\alpha_s(m_\tau^2)$. We conclude that
$\delta\alpha_s(m_\tau^2)\simeq\pm 0.03$ is a reasonable estimate of
the truncation error. This leads to
\begin{equation}\label{asval}
   \alpha_s(m_\tau^2) = 0.33\pm 0.03 \,, \qquad
   \alpha_s(m_Z^2) = 0.119\pm 0.004 \,.
\end{equation}
For the sake of completeness, we have translated our result into a
value of $\alpha_s$ at the mass of the $Z$ boson.

\begin{figure}[ht]
   \epsfxsize=9cm
   \centerline{\epsffile{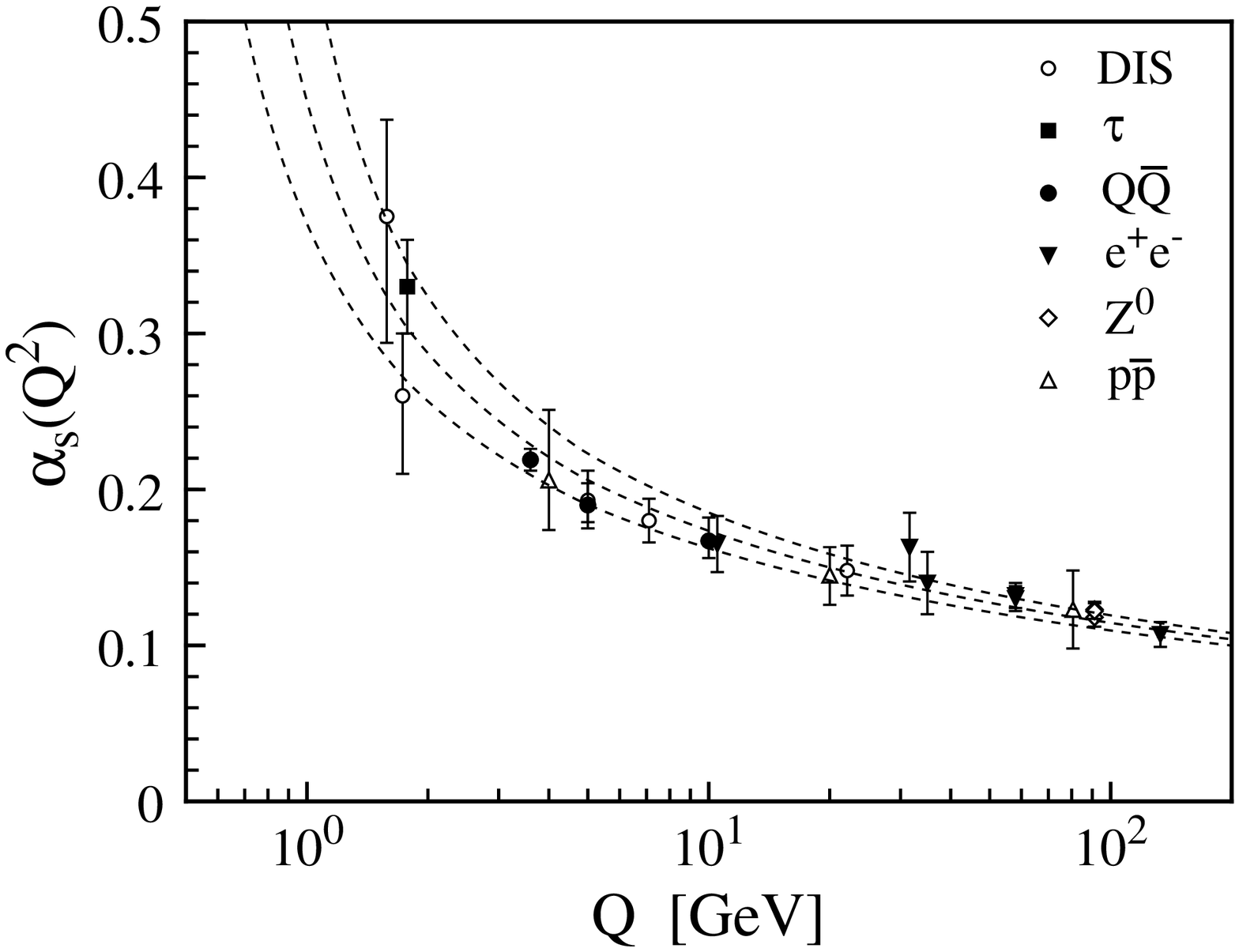}}
\fcaption{\label{fig:as_compile}
Compilation of $\alpha_s$ measurements. The curves correspond to the
QCD prediction for the running coupling constant for
$\alpha_s(m_Z^2)=0.116\pm 0.005$.}
\end{figure}

The analysis just described provides one of the best determination of
the QCD coupling constant in the low-energy region. The result
(\ref{asval}) is included in Fig.~\ref{fig:as_compile}, which shows a
collection of measurements of $\alpha_s$ performed at different
energy scales \cite{Webb}. Besides $\tau$ decays, low-energy ($Q\sim
1.6$--10~GeV) measurements come from deep-inelastic scattering and
$\Upsilon$ spectroscopy and decays. At higher energies ($Q\sim
30$--130~GeV), the most reliable determinations of $\alpha_s$ come
from measurements of the total cross section, jet rates and event
shapes in $e^+ e^-$, $p\bar p$ and $e p$ collisions. Taken all
together, these measurements provide clear evidence for the
``running'' of the effective coupling constant $\alpha_s(Q^2)$, which
in QCD is predicted to decrease with the momentum transfer. This
property of ``asymptotic freedom'' \cite{Gros} is one of the key
predictions of QCD. Formally, it is expressed by the fact that the
$\beta$-function is positive, where
\begin{eqnarray}\label{RGE}
   {{\rm d}\alpha_s(Q^2)\over{\rm d}\ln Q^2} &=& - \alpha_s(Q^2)\,
    \beta[\alpha_s(Q^2)] \,, \nonumber\\
   \beta(\alpha_s) &=& \beta_0\,{\alpha_s\over 4\pi}
    + \beta_1\,\bigg( {\alpha_s\over 4\pi} \bigg)^2
    + \beta_2\,\bigg( {\alpha_s\over 4\pi} \bigg)^3 + \dots \,,
\end{eqnarray}
and $\beta_0=9$, $\beta_1=64$ and $\beta_2=3863/6$ are the first
three expansion coefficients of the $\beta$-function, evaluated for
$n_f=3$ light quark flavours. (The value of $\beta_2$ is specific to
the $\overline{\mbox{\sc ms}}$ renormalization scheme.)

A test of the running of $\alpha_s$ by combining measurements
performed in many experiments operating at different energy scales
has the disadvantage of involving different experimental systematic
errors, as well as different levels of sophistication of the
theoretical calculations. Therefore, it is an appealing idea to
measure the scale dependence of the coupling constant in a single
experiment. This can be done in high-energy experiments at $p\bar p$
and $e p$ colliders, where a large range of $Q$ values can be probed
simultaneously \cite{Giel}. However, so far the precision obtained in
these measurements is rather low. In this talk, we propose a
high-precision test of the running of $\alpha_s$ in the low-energy
region ($0.85~\mbox{GeV}<Q<m_\tau$), using data obtained in a single
experiment \cite{Maria}. The value of $\alpha_s$ changes by about a
factor~2 in this energy range, which is equivalent to the variation
between 5 and 100~GeV.

\boldmath
\section{Extraction of $\alpha_s(Q^2)$ in $\tau$ decays}
\unboldmath

\begin{figure}[ht]
   \epsfxsize=8.5cm
   \centerline{\epsffile{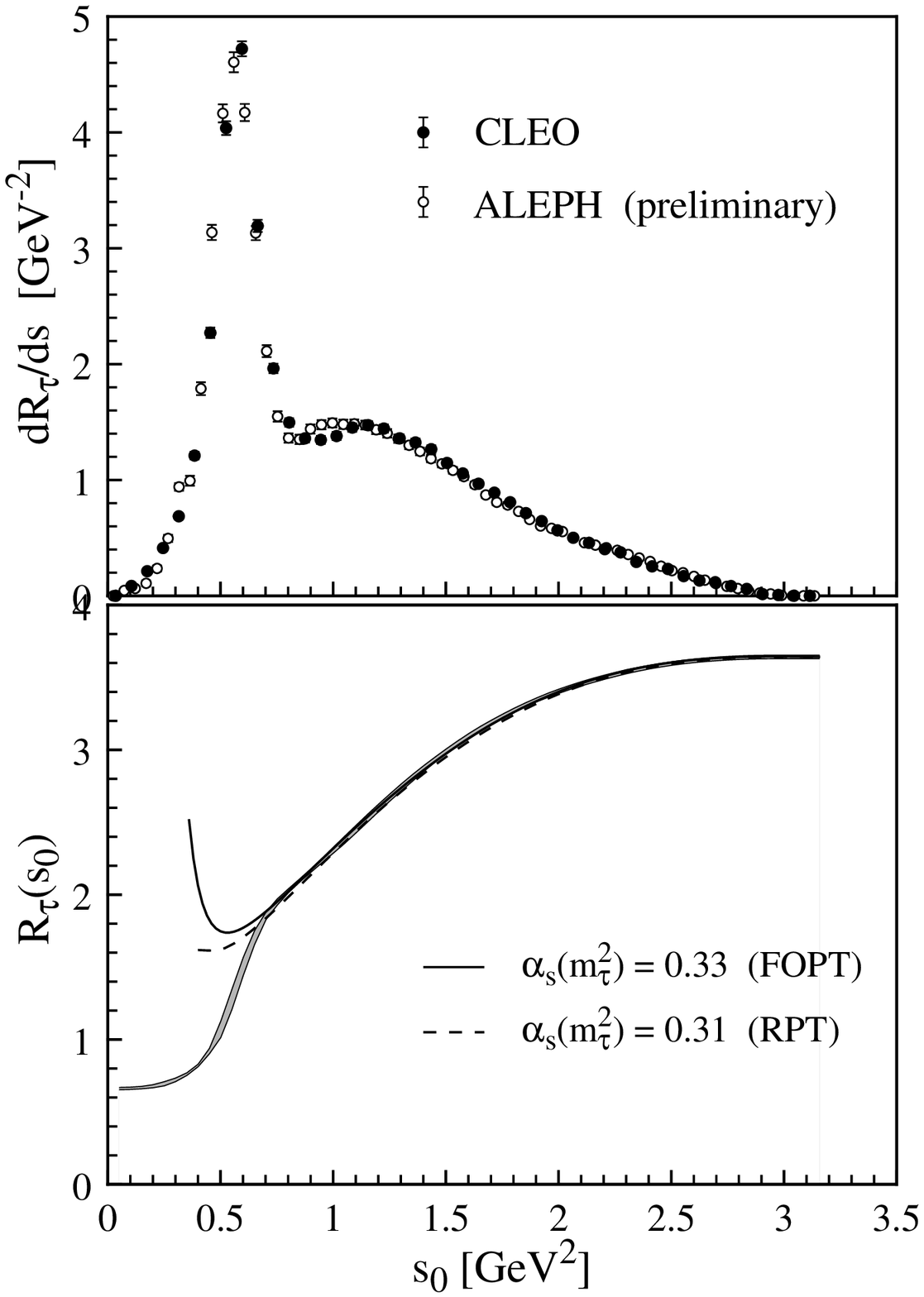}}
\fcaption{\label{fig:data}
Upper plot: The hadronic mass distribution ${\rm d}R_\tau/{\rm d}s$
in $\tau$ decays. Shown are the statistical errors after correcting
for detector effects. CLEO data from electron-tagged and muon-tagged
events have been combined. The ALEPH data are preliminary. Lower
plot: The integrated spectrum $R_\tau(s_0)$. The experimental result,
including statistical and systematic errors, is presented as a band.
The curves will be explained in Sect.~\protect\ref{sec:duality}.}
\end{figure}

We shall consider the $\tau$ decay rate into hadrons of invariant
mass squared smaller than $s_0$, normalized to the leptonic
decay rate:
\begin{equation}\label{Rs0def}
   R_\tau(s_0) = {\Gamma(\tau\to\nu_\tau + \mbox{hadrons};\,
   s_{\rm had} < s_0)\over\Gamma(\tau\to\nu_\tau\,e\,\bar\nu_e)}
   = \int\limits_0^{\displaystyle s_0}\!{\rm d}s\,
   {{\rm d}R_\tau(s)\over{\rm d}s} \,,
\end{equation}
where ${\rm d}R_\tau/{\rm d}s$ is the inclusive hadronic spectrum,
which has been measured by the CLEO and ALEPH Collaborations
\cite{CLEO,ALEPH}. To obtain ${\rm d}R_\tau/{\rm d}s$, we have
multiplied the normalized distributions by $R_\tau$. The result is
shown in the upper portion of Fig.~\ref{fig:data}. Not shown in the
figure is the contribution from $\tau\to h^-\nu_\tau$ with
$h^-=\pi^-$ or $K^-$, which has a branching ratio of $(11.77\pm
0.14)\%$ \cite{Callot}. Integrating these spectra over $s$ and
combining the results weighted by their statistical errors, we obtain
the distribution $R_\tau(s_0)$ shown in the lower portion of the
figure. Systematic errors have been estimated by taking the
difference between the CLEO and ALEPH data, and added in quadrature
with the statistical errors. Since the errors are strongly
correlated, the result is presented as a band.

Using the analyticity properties of QCD spectral functions, the
quantity $R_\tau(s_0)$ can be represented as a contour integral along
a circle of radius $|s|=s_0$ in the complex $s$-plane (for
simplicity, we quote the result in the chiral limit):
\begin{equation}\label{Rs0int}
   R_\tau(s_0) = {1\over 2\pi i}\,\oint\limits_{|s|=s_0}
   {{\rm d}s\over s}\,w\bigg({s_0\over m_\tau^2},{s\over m_\tau^2}
   \bigg)\,D(s) \,.
\end{equation}
Here
\begin{equation}
   w(x,y) = 2(x-y) - 2(x^3-y^3) + (x^4-y^4)
\end{equation}
is the phase-space function, and $D(s)$ is a current--current
correlation function, which contains all QCD dynamics. The
representation (\ref{Rs0int}) shows that $s_0$ is the only scale at
which QCD dynamics is probed; the $\tau$-lepton mass appears only in
the phase space. Provided that $s_0\gg\Lambda_{\rm QCD}^2$, the
correlation function $D(s)$ is needed at large momentum transfer
only, and the OPE can be employed to calculate $R_\tau(s_0)$ as a
function of $\alpha_s(s_0)$ and $x_0=s_0/m_\tau^2$:
\begin{equation}\label{Rs0OPE}
   R_\tau(s_0) = N_c\,\Big\{ r_{\rm pert}[\alpha_s(s_0),x_0]
   + r_{\rm power}(x_0) \Big\} \,.
\end{equation}

The perturbative contribution is given by
($a_0\equiv\alpha_s(s_0)/\pi$):
\begin{eqnarray}\label{rpert}
   r_{\rm pert}[\alpha_s(s_0),x_0]
   &=& (2 x_0 - 2 x_0^3 + x_0^4) \Big[ 1 + a_0 + 1.640 a_0^2
    - 10.28 a_0^3 + (K_4 - 156.0) a_0^4 \Big] \nonumber\\
   &&\mbox{}+ (2 x_0 - \textstyle\frac{2}{3} x_0^3
    + \textstyle\frac{1}{4} x_0^4)\,(2.25 a_0^2 + 11.38 a_0^3
    - 46.24 a_0^4) \nonumber\\
   &&\mbox{}+ (2 x_0 - \textstyle\frac{2}{9} x_0^3
    + \textstyle\frac{1}{16} x_0^4)\,(10.125 a_0^3 + 94.81 a_0^4)
    \nonumber\\
   &&\mbox{}+ (2 x_0 - \textstyle\frac{2}{27} x_0^3
    + \textstyle\frac{1}{64} x_0^4)\,68.34 a_0^4 + O(a_0^5) \,,
\end{eqnarray}
where $K_4$ is the five-loop coefficient in the Adler function, which
is currently not known exactly. In our analysis, we use the
estimate $K_4\simeq 27.5$ \cite{Kata} obtained using the methods of
Ref.~\cite{ECH}. As in the case of $R_\tau$, the truncation of the
perturbation series will turn out to be the main theoretical
uncertainty in our analysis. We estimate the importance of the
unknown higher-order contributions (of order $a_0^5$ and higher) by
resumming the renormalon-chain contributions to all orders in
perturbation theory, using the results of Ref.~\cite{MN}. We shall
compare fixed-order perturbation theory with this resummation and
take the difference as an estimate of the perturbative uncertainty.
This estimate of the truncation error is more conservative than that
obtained by dropping the last term in the series in (\ref{rpert}).

\begin{figure}[htb]
   \epsfxsize=9cm
   \centerline{\epsffile{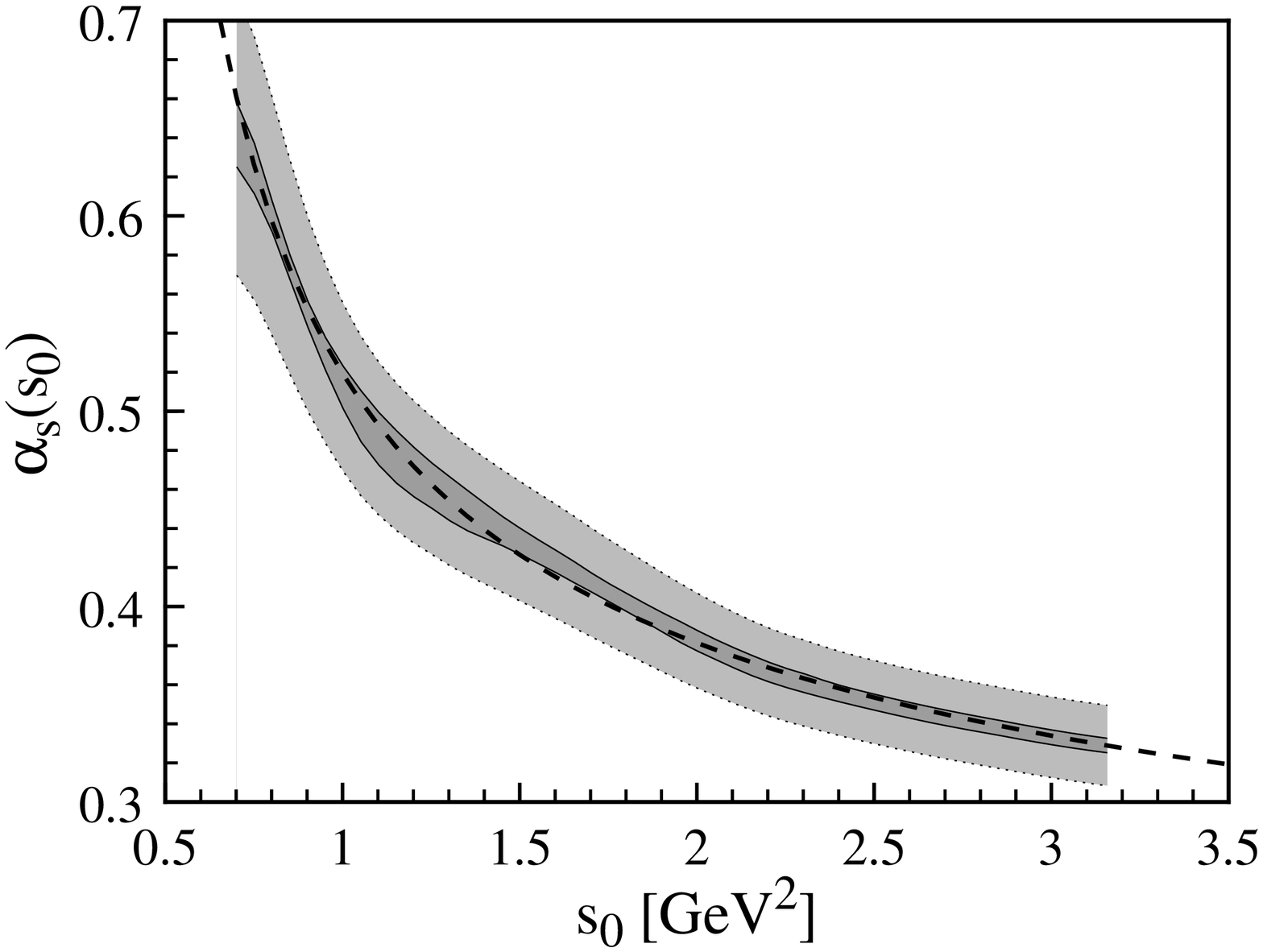}}
\fcaption{\label{fig:alphas}
Values of $\alpha_s(s_0)$ extracted from the data on $R_\tau(s_0)$.
The dark band represents the experimental errors, the light one the
sum of the experimental and theoretical errors. The errors are
strongly correlated. The dashed line shows the three-loop QCD
prediction for the running coupling constant.}
\end{figure}

The power corrections in (\ref{Rs0OPE}) are given by
\begin{equation}\label{rpower}
   r_{\rm power}(x_0) = -6\,|V_{us}|^2\,
   (1+x_0-x_0^2+\textstyle\frac{1}{3} x_0^3)\,
   \displaystyle{m_s^2\over m_\tau^2}
   + 32\pi^2\,{\langle m\bar\psi\psi\rangle\over m_\tau^4}
   - 2\,{\langle O_6\rangle\over m_\tau^6} + \dots \,.
\end{equation}
Note that, as a simple consequence of the representation
(\ref{Rs0int}), no inverse powers of $s_0$ appear in this
expression \cite{Maria}. This is true as long as the coefficients of
the power corrections to the correlation function $D(s)$ do not
contain logarithms of $s$. As a result, the OPE converges well down
to low scales $s_0$. For instance, we find $r_{\rm power}=-(1.4\pm
0.5)\%$ at $s_0=m_\tau^2$, and $r_{\rm power}=-(1.5\pm 0.5)\%$ at
$s_0=1$~GeV$^2$. The break-down of the OPE (see
Sect.~\ref{sec:duality} below) will thus not be driven by a blow-up
of the series of power corrections. Another important feature of
(\ref{rpower}) is that the terms involving the vacuum condensates are
independent of $s_0$. Hence, the uncertainties in the values of the
condensates do not affect the $s_0$ dependence of $R_\tau(s_0)$,
which will be used to study the running of $\alpha_s(s_0)$.

{}From the measurement of the quantity $R_\tau(s_0)$ shown in
Fig.~\ref{fig:data}, we extract $\alpha_s(s_0)$ as a function of
$s_0$ by fitting to the data the theoretical prediction obtained
using fixed-order perturbation theory. The result, including
experimental errors only, is represented by the dark band in
Fig.~\ref{fig:alphas}. Theoretical uncertainties arise from the
truncation of the perturbation series and from the uncertainty in the
values of the nonperturbative parameters. They affect the overall
scale of the $\alpha_s$ values (by about 8--10\%), but have very
little effect on the evolution of the coupling constant. The sum of
the experimental and theoretical errors is represented by the light
band. The dashed curve shows the QCD predictions for $\alpha_s(s_0)$
obtained at three-loop order, normalized to the central value of the
data at $s_0=m_\tau^2$. The observed scale dependence of the running
coupling constant is in excellent agreement with the QCD prediction.

\begin{figure}[ht]
   \epsfxsize=10cm
   \centerline{\epsffile{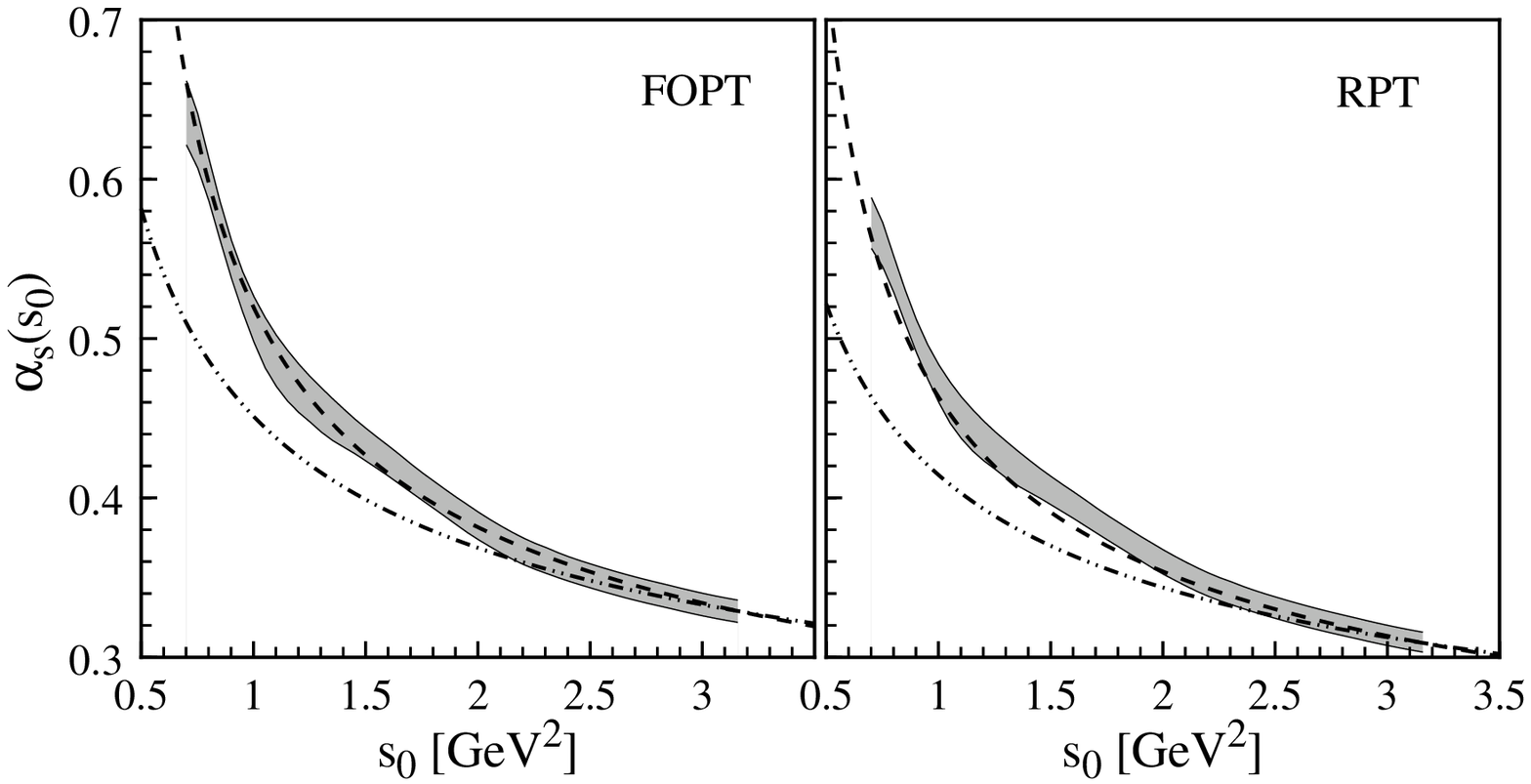}}
\fcaption{\label{fig:alphas2}
Values of $\alpha_s(s_0)$ extracted from the data on $R_\tau(s_0)$
using fixed-order (FOPT) and resummed perturbative theory (RPT). The
dashed lines show the QCD prediction obtained using the three-loop
$\beta$-function. The dash-dotted lines refer to the one-loop
$\beta$-function.}
\end{figure}

As mentioned above, the main theoretical uncertainty comes from the
truncation of the perturbation series. This is illustrated in
Fig.~\ref{fig:alphas2}, where the evolution of $\alpha_s(s_0)$ as a
function of $s_0$ is shown separately for fixed-order and resummed
perturbation theory. The curves show the QCD predictions for the
running coupling constant obtained at one- and three-loop order,
normalized to the data at $s_0=m_\tau^2$. It is seen that
higher-order corrections effectively renormalize the overall scale of
the $\alpha_s$ values (i.e.\ the $\Lambda_{\rm QCD}$ parameter). For
instance, the value of $\alpha_s$ at $s_0=m_\tau^2$ changes from 0.33
(fixed-order) to 0.31 (resummed). The difference between the two
results for $\alpha_s(s_0)$ has been used to estimate the truncation
error.

\begin{figure}[htb]
   \epsfxsize=10cm
   \centerline{\epsffile{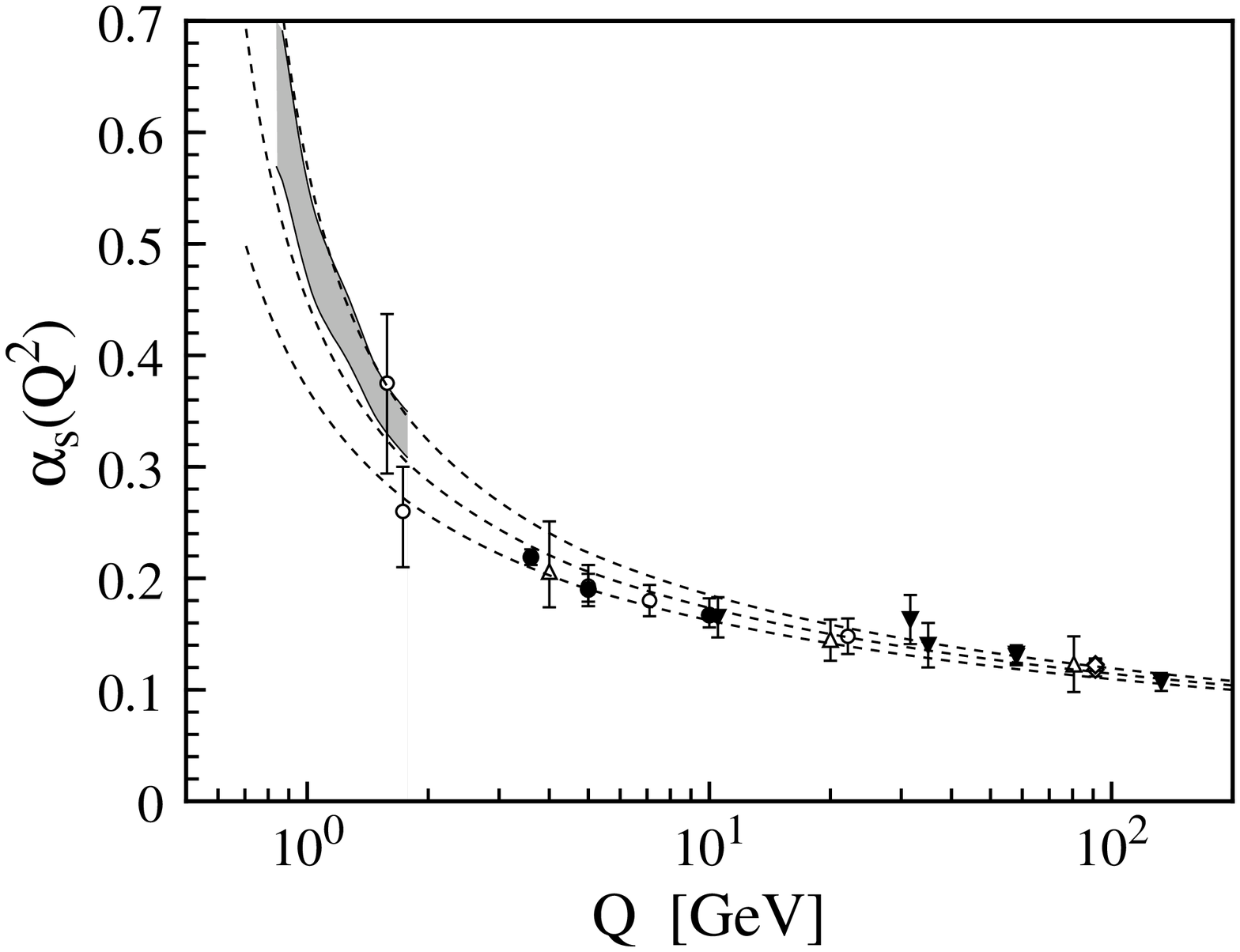}}
\fcaption{\label{fig:run}
Compilation of $\alpha_s$ measurements including our result obtained
from the analysis of hadronic $\tau$ decays. The curves correspond to
the QCD prediction for the running coupling constant for
$\alpha_s(m_Z^2)=0.121$, 0.116, 0.111 (top to bottom).}
\end{figure}

Fig.~\ref{fig:run} shows our result combined with the other
measurements of $\alpha_s$ collected in Fig.~\ref{fig:as_compile}. We
have replaced the data point at $Q=m_\tau$ by the band shown in
Fig.~\ref{fig:alphas}, which extends to much lower values of $Q$.
This figure demonstrates nicely the main features of our approach: it
extends the range of $\alpha_s$ values accessible to experiments,
thus allowing a measurement of the strong coupling constant at scales
lower than the lowest ones attainable before. Moreover, it provides a
test of the QCD evolution of $\alpha_s$ with higher precision than
all other single measurements of the running to date.

\section{Break-down of the OPE and quark--hadron duality}
\label{sec:duality}

An important question which we have to address is to determine the
lowest value of $s_0$ for which our analysis can be trusted. In other
words, at which point do we expect the OPE to break down? To answer
this question, it is important to realize that we are applying the
OPE in the physical region (i.e.\ the region of time-like momenta),
where QCD cannot be used to calculate correlation functions such as
$D(s)$. The reason why we trust the calculation of $R_\tau(s_0)$ is
that to perform the contour integral in (\ref{Rs0int}) requires
knowledge of the correlation function for large (complex) momenta
only. Moreover, the integrand vanishes for $s=s_0$, where the contour
touches the branch cut of $D(s)$; hence, the main contributions come
from regions far away from the singularities, where the OPE can be
applied. Another way to say this is that in the calculation of
$R_\tau(s_0)$ we assume quark--hadron duality, which is the
hypothesis that QCD can be employed to calculate physical decay rates
if they are ``smeared'' over a sufficiently wide energy interval
\cite{PQW}. In the present case, this smearing is provided by the
integration over the range $0<s<s_0$ in (\ref{Rs0def}). The question
of how accurate the duality assumption is and for what values of
$s_0$ it applies is, however, a phenomenological one. Despite of some
interesting new ideas \cite{Shifm}, it cannot be answered yet from
theoretical grounds.

To test the assumption of duality, we compare the data for the
quantity $R_\tau(s_0)$ with the theoretical predictions obtained from
the OPE, using both fixed-order and resummed perturbation theory. The
results are shown by the two curves in the lower portion of
Fig.~\ref{fig:data}. In obtaining these curves, we have adjusted the
value of $\alpha_s(m_\tau^2)$ so as to fit the data at
$s_0=m_\tau^2$. The value of $\alpha_s(s_0)$ is then obtained from
the solution of the renormalization-group equation (\ref{RGE}).
Theoretical uncertainties have little influence on the $s_0$
dependence of $R_\tau(s_0)$. For the perturbative part of the
calculation, this is apparent from the good agreement of the two
theoretical curves in Fig.~\ref{fig:data}, which refer to values of
$\alpha_s(m_\tau^2)$ that differ by 9\%. Hence, the $s_0$ dependence
of $R_\tau(s_0)$ is predicted essentially without any free
parameters, and the comparison of the data with the theoretical
predictions provides a direct test of quark--hadron duality.

We find excellent agreement over the range
$0.7~\mbox{GeV}^2<s_0<m_\tau^2$, indicating that in $\tau$ decays
duality holds as soon as the integral over the hadronic mass
distribution includes the $\rho$ resonance peak. This justifies {\em
a posteriori\/} our choice of the energy interval in the previous
section. It is remarkable that, once $s_0$ exceeds the value of
0.7~GeV$^2$, the onset of duality happens almost instantaneously.
Since the $\rho$ meson is such a prominent resonance, this is the
best possible scenario that could be expected. The small oscillation
of the experimental band around the theoretical curve, which could be
due to some deviations from duality in the region of the $a_1$
resonance, are not significant given the precision of the data. Even
if such oscillations will be confirmed in further analyses based on
more precise data, they will clearly not put a severe limitation on
the applicability of our method.

\boldmath
\section{Measurement of the $\beta$-function}
\unboldmath

\begin{figure}
   \epsfxsize=9cm
   \centerline{\epsffile{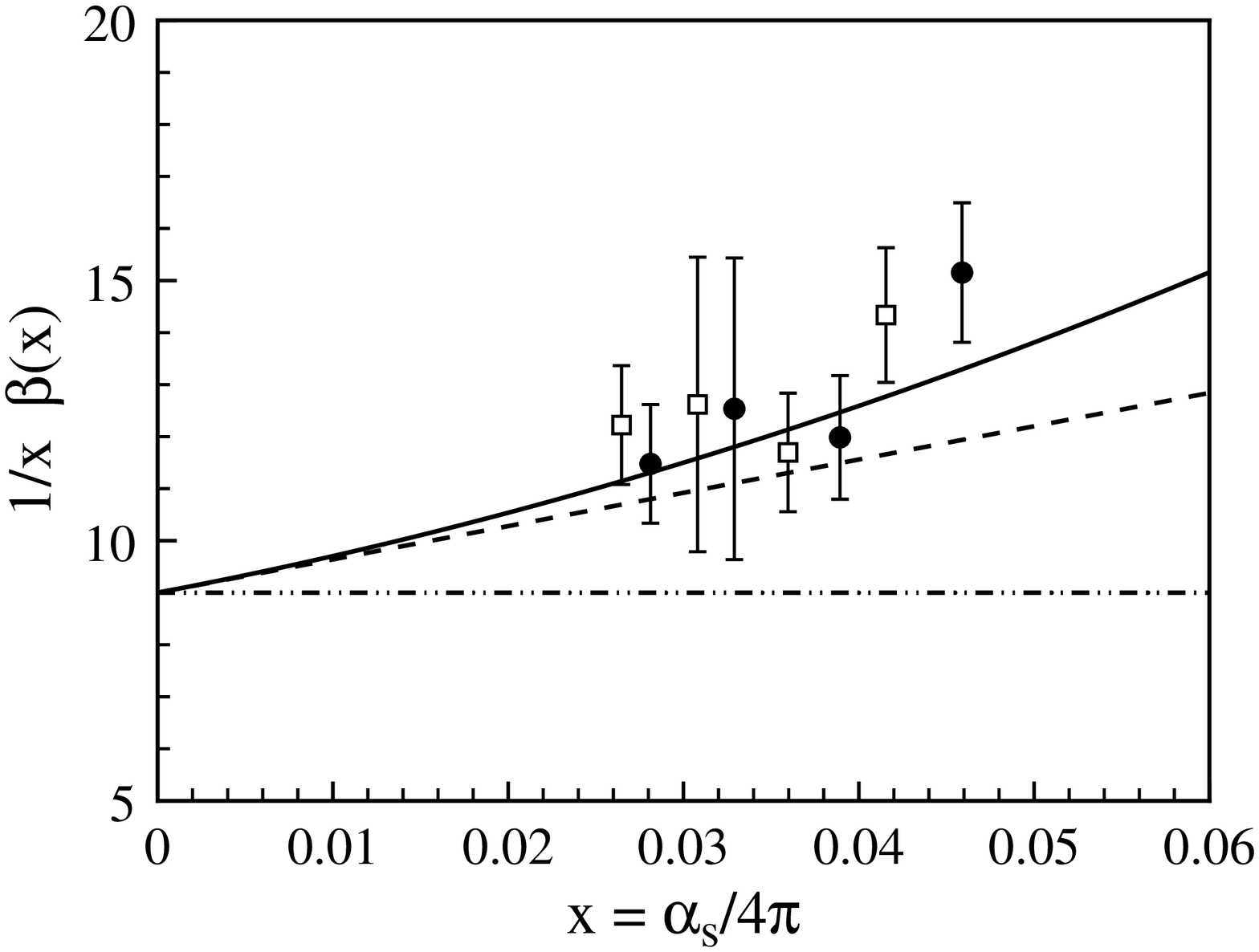}}
\fcaption{\label{fig:betafun}
Experimental determination of the $\beta$-function. The circles are
obtained using fixed-order perturbation theory, the squares refer to
resummed perturbation theory. The curves show the QCD
$\beta$-function at one-loop (dash-dotted), two-loop (dashed) and
three-loop (solid) order.}
\end{figure}

To quantify the agreement between the data and the QCD prediction for
the running coupling constant exhibited in Figs.~\ref{fig:alphas} and
\ref{fig:alphas2}, we extract from the data the $\beta$-function
defined in (\ref{RGE}) and compare the result to the prediction of
QCD perturbation theory. Introducing the variable
$x=\alpha_s(s_0)/4\pi$, we have
\begin{equation}\label{beta}
   - {4\pi\over\alpha_s^2(s_0)}\,
   {{\rm d}\alpha_s(s_0)\over{\rm d}\ln s_0} = {\beta(x)\over x}
   = \beta_0 + \beta_1 x + \beta_2 x^2 + \dots \,.
\end{equation}
We approximate the derivative ${\rm d}\alpha_s/{\rm d}\ln s_0$ by a
ratio of differences, $\Delta\alpha_s/\Delta\ln s_0$, for a selected
set of $s_0$ values chosen such that the differences $\Delta\alpha_s$
are large enough to be significant given the errors in the
measurement. For $\alpha_s(s_0)$ in (\ref{beta}) we take the central
value of each interval. We use the following $s_0$ values: 0.75,
0.95, 1.35, 2.06, and 3.16~GeV$^2$, corresponding to four intervals
of increasing width $\Delta\ln s_0$, but constant
$\Delta\alpha_s\simeq 0.075$. The results are shown in
Fig.~\ref{fig:betafun}. The circles are obtained using fixed-order
perturbation theory, while the squares refer to resummed perturbation
theory. As expected, the two methods give very similar results for
the running of the coupling constant. The estimate of the errors
includes the theoretical uncertainties, the error due to the choice
of finite intervals in $\alpha_s$, and the experimental errors, which
in this case are the dominant ones. The curves in
Fig.~\ref{fig:betafun} show the QCD $\beta$-function at one-, two-
and three-loop order in perturbation theory. The data provide clear
evidence for the running of the coupling constant. Moreover, they
prefer a running that is stronger than predicted at one-loop order.
Indeed, between the three curves, the best description of the data is
provided by the three-loop prediction. Performing a fit with the
three-loop $\beta$-function, where $\beta_0=9$ and $\beta_1=64$ are
kept fixed but the three-loop coefficient $\beta_2$ is treated as a
parameter, we find $\beta_2^{\rm exp}/\beta_2^{\rm th}=1.6\pm 0.7$
using fixed-order perturbation theory, and $1.8\pm 0.8$ using
resummed perturbation theory.

We believe that such an experimental determination of the
$\beta$-function beyond the leading order can at present be done only
in $\tau$ decays. (A high-precision measurement of $R_{e^+ e^-}(s)$
in the region below the charmonium resonances would provide an
alternative place for such a study.) At higher energies, the value of
$\alpha_s$ is too small to distinguish between the three curves in
Fig.~\ref{fig:betafun}; measurements in the region $Q\sim 100$~GeV,
for instance, correspond to values $x\sim 0.01$.

\section{Conclusions}

We have presented a method to measure the running coupling constant
$\alpha_s(Q^2)$ in the low-energy region $0.85~\mbox{GeV}<Q<m_\tau$,
using $\tau$-decay data obtained in a single experiment. At
$Q=m_\tau$, we obtain $\alpha_s(m_\tau^2)=0.33\pm 0.03$,
corresponding to the rather precise value $\alpha_s(m_Z^2)=0.119\pm
0.004$. Our method provides a test of the scale dependence of the
coupling constant in a region where this effect is most pronounced.
The theoretical analysis is based on the OPE and the assumption of
quark--hadron duality. We have tested this assumption and find that
it holds provided the $\tau$ decay rate is integrated over an energy
interval large enough to include the $\rho$ resonance peak. Our
analysis provides a test of QCD at scales comparable with the lowest
ones attainable before ($Q\simeq 1.6$~GeV in deep-inelastic
scattering), and with higher precision than all other single
measurements of the running to date. We have extracted for the first
time the $\beta$-function from data and find that it is in good
agreement with the three-loop prediction of QCD.


\begin{thebibliography}{99}

\bibitem {SVZ}
M.A. Shifman, A.I. Vainsthein and V.I. Zakharov, Nucl.\ Phys.\ B
{\bf 147}, 385 and 448 (1979).

\bibitem {Rtau1}
E. Braaten, S. Narison and A. Pich, Nucl.\ Phys.\ B {\bf 373}, 581
(1992).

\bibitem {Callot}
G. Rahal-Callot, to appear in: Proc.\ Int.\ Europhysics Conf.\ on
High Energy Physics, Brussels, Belgium, September 1995.

\bibitem {Wash}
S.R. Wasserbaech, to appear in: Proc.\ Workshop on the Tau--Charm
Factory, Argonne, Illinois, June 1995.

\bibitem {Alta}
G. Altarelli, P. Nason and G. Ridolfi, Z.\ Phys.\ C {\bf 68}, 257
(1995).

\bibitem {LP}
F. Le Diberder and A. Pich, Phys.\ Lett.\ B {\bf 286}, 147 (1992);
{\bf 289}, 165 (1992).

\bibitem {tHof}
G. 't Hooft, in: Proc.\ 15th Int.\ School of Subnuclear Physics,
Erice, Sicily, 1977, ed.\ A.~Zichichi (Plenum Press, New York, 1979),
p.~943;
B. Lautrup, Phys.\ Lett.\ B {\bf 69}, 109 (1977);
G. Parisi, Phys.\ Lett.\ B {\bf 76}, 65 (1978);
Nucl.\ Phys.\ B {\bf 150}, 163 (1979);
F. David, Nucl.\ Phys.\ B {\bf 234}, 237 (1984); {\bf 263}, 637
(1986);
A.H. Mueller, Nucl.\ Phys.\ B {\bf 250}, 327 (1985);
V.I. Zakharov, Nucl.\ Phys.\ B {\bf 385}, 452 (1992);
M. Beneke and V.I. Zakharov, Phys.\ Rev.\ Lett.\ {\bf 69}, 2472
(1992);
D. Broadhurst, Z.\ Phys.\ C {\bf 58}, 339 (1993).

\bibitem {BBB}
P. Ball, M. Beneke and V.M. Braun, Nucl.\ Phys.\ B {\bf 452}, 563
(1995);
C.N. Lovett-Turner and C.J. Maxwell, Nucl.\ Phys.\ B {\bf 452}, 188
(1995).

\bibitem {MN}
M. Neubert, Nucl.\ Phys.\ B {\bf 463}, 511 (1996).

\bibitem {Webb}
B.R. Webber, in: Proc.\ 27th Int.\ Conf.\ on High Energy Physics,
Glasgow, Scotland, July 1994, eds.\ P.J.~Bussey and I.G.~Knowles (IOP
Publ., Bristol, 1995), Vol.~1, p.~213;
R.K.~Ellis and B.R.~Webber, private communication.

\bibitem {Gros}
D.J. Gross and F. Wilczek, Phys.\ Rev.\ Lett.\ {\bf 30}, 1343
(1973);
H.D. Politzer, Phys.\ Rev.\ Lett.\ {\bf 30}, 1346 (1973).

\bibitem {Giel}
M. Derrick et al.\ (ZEUS Collaboration), Phys.\ Lett.\ B {\bf 363},
201 (1995);
W.T. Giele, E.W.N. Glover and J. Yu, FERMILAB-Pub-95/127-T (1995)
[hep-ph/9506442].

\bibitem {Maria}
M. Girone and M. Neubert, Phys.\ Rev.\ Lett.\ {\bf 76}, 3061 (1996).

\bibitem {CLEO}
T. Coan et al.\ (CLEO Collaboration), Phys.\ Lett.\ B {\bf 356}, 580
(1995).

\bibitem {ALEPH}
L. Duflot, in: Proc.\ 3rd Workshop on Tau Lepton Physics, ed.\
L.~Rolandi, Nucl.\ Phys.\ B (Proc.\ Suppl.) {\bf 40}, 37 (1995).

\bibitem {Kata}
A.L. Kataev and V.V. Starshenko, Mod.\ Phys.\ Lett.\ A {\bf 10}, 235
(1995).

\bibitem {ECH}
P.M. Stevenson, Phys.\ Rev.\ D {\bf 23}, 2916 (1981);
G. Grunberg, Phys.\ Lett.\ B {\bf 221}, 70 (1980); Phys.\ Rev.\ D
{\bf 29}, 2315 (1984).

\bibitem {PQW}
E.C. Poggio, H.R. Quinn and S. Weinberg, Phys.\ Rev.\ D {\bf 13},
1958 (1976).

\bibitem {Shifm}
M. Shifman, TPI-MINN-95/15-T (1995) [hep-ph/9505289], to appear in:
Proc.\ Joint Meeting of the Int.\ Symp.\ on Particles, Strings and
Cosmology \& 19th Johns Hopkins Workshop on Current Problems in
Particle Theory, Baltimore, Maryland, March 1995.

\end{thebibliography}
\end{document}